\documentclass[10pt,journal,compsoc]{IEEEtran}

\ifCLASSOPTIONcompsoc
   \usepackage[nocompress]{cite}
\else
    \usepackage{cite}
\fi

\ifCLASSINFOpdf
 
\else
 
\fi

\usepackage{graphicx}
\usepackage{xcolor}
\usepackage{tabularx}

\usepackage{authblk}

\begin{document}

\title{Impacts and Integration of Remote-First Working Environments}

 

\author{\IEEEauthorblockN{Christopher Atti\IEEEauthorrefmark{1}, Cliff Cross\IEEEauthorrefmark{1}, Ahmet Bugra Dogan\IEEEauthorrefmark{1},Christopher Hubbard \IEEEauthorrefmark{1}, Cameron Page \IEEEauthorrefmark{1}, Stephen Montague  \IEEEauthorrefmark{1},
Elnaz Rabieinejad\IEEEauthorrefmark{2}	}
	\IEEEauthorblockA{
	 \\
		\IEEEauthorrefmark{1}College of Computing and Software Engineering,\\ Kennesaw State University, Marietta, GA USA
 \\catti@students.kennesaw.edu, ccross6@students.kennesaw.edu, adogan2@students.kennesaw.edu,
chubba29@students.kennesaw.edu,
smontagu@students.kennesaw.edu,
 \\
					\IEEEauthorrefmark{2}Cyber Science Lab, School of Computer Science, University of Guelph,	Ontario, Canada \\  erabiein@uoguelph.ca  \\
							}}




\IEEEtitleabstractindextext{%
\begin{abstract}
\textcolor{black}{Due to the Covid-19 pandemic in 2020 or other business decisions, remote work is becoming increasingly popular. “Remote first” working environments exist within companies where most employees work remotely. This paper takes a deep dive into the remote-first mentality. It investigates its effects on employees at varying stages in their careers, day-to-day productivity, and working relationships with team members. We found that the remote-first mentality most impacts seasoned employees and managers, potentially due to trouble adjusting to a new way of working compared to the rest of their careers and the “always on” mentality associated with working from home. 
Regarding productivity, we found that while software development productivity appears unimpacted, the effectiveness of communication and employee wellbeing saw declines which are generally associated with lowered productivity. Finally, we looked closer at the communication side of things and how remote work impacts relationship building. We found that the most significant impacts on relationship building centered around “trust” and “credibility” being harder to build due to a lack of non-verbal cues during social interactions.  
 }
\end{abstract}

\begin{IEEEkeywords}
Remote-first, Work from home, Virtual relationship building, Remote communication, Covid-19 impacts, Challenges of remote work, Remote work productivity, Virtual collaboration.
\end{IEEEkeywords}}

\maketitle

\IEEEdisplaynontitleabstractindextext

\IEEEpeerreviewmaketitle

\section{Introduction}

In today’s world, many companies face the dilemma of allowing or becoming a remote-first work environment. Companies have allowed people to work from home for quite some time now. Most likely, it was one, maybe two, days a week. Therefore, “work from home” is not exactly a new phenomenon. On the other hand, one could argue that working in an office is the new ideology \cite{a1,a2}. According to WeWorkRemotely.com, “Before the Industrial Revolution, everyone worked out of their homes \cite{1}. Skilled blacksmiths, carpenters, leather workers, and potters each set up shop at their residence and sold their goods from there.” The Industrial Revolution was when big machines and the need to mass-produce forced people to go into warehouse-type buildings where they could work in such fashion. Of course, some still require large machinery and mass production, but most desk jobs do not require employees to be in a centralized building to get their work done productively. In fact, “’43 percent of employed Americans said they spent at least some time working remotely’ according to one Gallup study” \cite{1}.  
The article posted this quote was written in 2017, long before COVID-19 forced every non-essential worker to work remotely \cite{a3}. Then, companies had to adapt quickly to having almost all their employees in remote positions, which has become a revolution. However, as we head towards a light at the end of the tunnel regarding the pandemic that forced millions of people to work remotely, companies require their employees to return to the office at total capacity. A new term is being coined in response to employees who are being forced back into the office. It is known as “The Great Resignation.” According to Forbes, “Dubbed ‘The Great Resignation,’ workplaces are seeing a trend in employees quitting their jobs. A record four million workers called it quits in April alone, according to the Labor Department. Pandemic burnout, a collective reassessment about priorities and what matters most, a labor shortage and more companies calling workers back into the office maybe some of the reasons behind this trend”  \cite{2}. A direct result of people being asked to return to the office is causing “a record four million workers” to quit their jobs in search of something that allows for more flexibility \cite{a4}. Now, people want to work from home or have the flexibility to choose a remote-first environment, but is that the best-case scenario for all employees or companies? In this paper,  firstly, we look to determine the effects remote-first work has on employees at different career stages. Then, we discuss an employee’s productivity level in the office versus at home and the process and difficulty it takes in getting a remote-first team to gel and work together well. Finally, we examine the benefit of Artificial intelligence (AI) in the remote working environment.

\section{CHALLENGES OF REMOTE WORK AT VARYING CAREER STAGES}\label{ExSurv}

\subsection{Seasoned Versus New Employees }\label{ExSurv}

While working remotely, many employees face new challenges that you may not be subject to in an office setting \cite{a5,a6}. In recent years, seasoned employees have adjusted to becoming fully remote while newer employees most likely have been hired on for a start to work full remote; both have their challenges.   
Seasoned employees could have a more challenging time adjusting to working from home  \cite{3}. Real-world distractions around them that they did not have before, such as housework, pets, tv, family, and outside noise, could all play a part in entirely focusing on the job. This is because this is a newer problem that these workers face. Some remote employees find it helpful to have a dedicated home office where they go to do work. This can help those feeling like they never get away from work getaway. Once they leave their office, they leave work there. Seasoned employees that have been used to going to the office sometimes can have difficulty adjusting to working remotely. The transition from the office environment where they have gotten used to interactions daily with people can also be challenging \cite{4}. While working remotely, many of the day-to-day interactions will be conducted virtually. Some remote employees also feel distanced from their team because they do not see them in person. If available, these employees may enjoy a hybrid style work environment where they can work in the office one or two days a week.   
On the flip side, newly hired employees were most likely already engaged to be remote or have a hybrid role at home and in the office. However, these new employees may not understand the company rules and standards needed to do the job. According to  \cite{4}, One of the biggest challenges would most likely fall under communication issues. While working remotely, one has to be very connected to those they work with, especially if they have a team-oriented role. Being dialed into what one’s team is doing is essential for completing tasks correctly. Sometimes not having a direct face-to-face interaction about a project can limit one’s understanding of what it takes to complete and not ask the right questions to get the correct answers  \cite{5}. That could mean going back and forth to fix mistakes along the way.

\subsection{Seasoned Versus New Employees }

Managers face different types of challenges when it comes to their employees. One thing that must be done by managers that have remote teams is to make sure they are communicating consistently with them about projects and what is expected of them \cite{6}. In addition, managers must keep their experienced employees in the loop of expectations. If their employees are experienced, there will be less concern about getting their job done than losing productivity or having other distractions happen \cite{a7,a8}.   
Managers tend to have an “always on” approach to work, especially while being remote. After hours, remote workers may answer emails or calls more than if one left the office and went home. That feeling of not disconnecting can be a challenge for many managers because they feel they can get more completed or try to answer that last email or finish up that previous request \cite{7}. However, disconnecting can be very important to unwind and keep that work and life balance separate.

\section{EMPLOYEE PRODUCTIVITY IN OFFICE AND REMOTE}
\subsection{Measures of Remote Work Productivity  }

Productivity can be measured by different metrics, both quantitative and qualitative. Typical measurements are deployment frequency, review to merge time, TLOC, etc. The degree to which a remote employee is productive is more scrutinized than employees in the office and more reliant on complex data. Observing an in-office employee toiling away can lend more credence to visible productivity versus seeing an “online” status indicator. As noted by “Darja, Anastasiia, Nils, Efi, Eriks, and Marte, cited from Smite, Russo, and Bezerra. “The analysis of all GitHub projects (open source and corporate) concludes that the developer activity in terms of the number of pushes, pull requests, code reviews, and commented issues remained similar or slightly increased compared to the pre-pandemic year. This finding is consonant with numerous other studies that conclude that software companies have nothing to worry about since working from home is per se not a significant challenge for software engineers” \cite{8}. They express how the value of activity metrics cannot answer questions regarding factors such as how to measure periods of inactivity between actions and engagement and concentration periods \cite{a9,a10}.

\subsection{	Reported Productivity}
Self-reported productivity among engineers who work from home differs based on personal preference and everyone’s barriers and drivers. Darja, Anastasiia, nils, Efi, Eriks, and Marte, analyze six corporate surveys conducted in four Scandinavian companies to uncover “emotional issues” and factors stemming from perceived isolation and communication style/preference are barriers for those employees who report decreased productivity. Better work-life balance, “flexibility and planning the work hours,” and the removal of commuting and pop-up conversations were among the most prevalent reported benefits, and drivers of working from home \cite{8}. The above group explained how personal biases could further skew the reliability of remote work survey data from a top-down organizational level to individual reflections.

\subsection{	Productivity in an Agile Environment}
Issues in standard work methodologies like Scrum become compounded when workers are remote. Mak and Krutchen study the problems with productivity via task coordination. "Agile techniques such as features or story cards tend not to adapt very well \cite{9}". The impact of Techniques that utilize physical presence, such as "Dot Voting" and "Fist of Five," are minimized. The names of these activities, ex. "StandUp" becomes more labels than actions, but the scrum guide does not explicitly state that a team must be positioned in the same place.  
Along with difficulties in team dynamics, there is the perception that trust suffers under a distributed agile team. In establishing an empirically developed framework for building trust in remote agile teams: Tyagi, Sibal, and Suri report, "Trust is one of the key factors that influences team performance and acts as a foundation for effective teamwork" in agile teams \cite{10}. The authors conceive a framework that assists and supports agile teams that can benefit from trust-building techniques and exercises.  
Communication challenges can become the bottleneck of an effective agile environment in remote development environments. Bundhun and Roopesh create a communication framework to address productivity attributes such as" differences in geographical locations and trust and cultural issues between team members and poor coordination and communication," as observed by a Mauritian-based software company \cite{11}. In leveraging agreed-upon communication standards, these authors could provide a framework and survey results to accomplish adaptable guidelines.

\subsection{	Employee Wellbeing Related to Productivity}
It is generally accepted that higher well-being is related to a more performant employee. However, general well-being can be attributed to many variables, including timeline pressures, work-from-home conflicts, and work-family conflicts. This is demonstrated by Darouei and Pluut, who surveyed 34 professional workers to test their five hypotheses. \\
"Hypothesis 1. Working from home (compared with the office) will be negatively associated with work-family conflict. \\
Hypothesis 2. Within individuals, time pressure mediates the negative relationship between working from home and work-family conflict experienced at home. \\
Hypothesis 3. Within individuals, work-family conflict experienced at home in the evening is positively related to emotional exhaustion the next morning. \\
Hypothesis 4. Within individuals, work-family conflict experienced at home in the evening is negatively related to work engagement the next morning. \\
Hypothesis 5a. Within individuals, work-family conflict experienced at home in the evening is positively related to negatively affecting the organization the next morning. \\
Hypothesis 5b. Within individuals, work-family conflict experienced at home in the evening is negatively related to positive affect towards the organization the next morning". The research survey results generally supported the hypotheses presented, with either a negative or positive outcome affecting performance directly \cite{12}.

\section{	DIFFICULTIES IN TEAMWORK}
Semi-related to productivity is a team’s ability to “gel” and foster healthy personal and working relationships \cite{a11}. Of course, many factors go into team dynamics and working relationships. However, for the sake of analysis and keeping separate from the “productivity” discussion, we can narrow this down to the ability to get to know your team and perform collaborative software development within a given toolset.

\subsection{	Difficulties in Getting to Know Your Team}
As Wojahn, Taylor, and Blicharz identified, remote-first employees, interact with colleagues from all over the country and the globe. They do this using a combination of voice, text, and video communication, with the primary means of communication being text or voice only. This leads to complications when attempting to gauge others’ reactions to us due to the lack of visual clues such as body language, facial expressions, and other non-verbal clues. Even more difficult than maintaining existing relationships and communication with colleagues when a new team comes together remotely and meets each other for the first time \cite{13}). The complications with non-visual communication are exacerbated when team members have no prior history of working experience together. In their paper, Moster, Ford, and Rodeghero identify that trust-building is one of the most challenging parts of remote work. Communication is vital for building trust, and “communication for virtual teams is often less frequent than in-person teams.” To combat complications due to inconsistent communication and lack of initial faith, ad-hoc personal conversations and video conferencing early in team formation can lessen the impact of virtual teamwork \cite{14}.  
In a remote-first environment, the primary means of communication will be digital, whether email, instant message, voice/video chat, or some other form of virtual communication that is not in-person \cite{a12}. According to Yang, the previously mentioned communication mediums fall under the “computer mediated communication” category and are supported by many tools to help facilitate relationship building \cite{a13}. Trust is difficult to build virtually, but “moreover, when all social interactions are mediated by technology, people’s impressions and perception of each other can be affected by the mediated social cues transmitted by different media. For example, the profile images people use, the description of one’s self-introduction, and visible online activities can influence how people perceive one another and their willingness to form further relationships to various extents.” He conducted multiple experiments involving virtual interactions and found that the lack of non-verbal cues can impact relationships with team members. However, poorly chosen ambient cues also add a factor to non-verbal cues. Virtual backgrounds or avatars deemed “unprofessional” were found to impact perceptions of interactions and employee credibility negatively \cite{15}.  

\subsection{	Difficulties in Collaborative Software Development}
The most critical factor in remote-first work is communication, but communication alone will not solve all collaboration issues. Gupta and Fernandez group collaboration in software engineering into four categories: model-based collaboration to reduce ambiguity and increase error detection, process-centric tools to help define and facilitate the software process, awareness tools to enable team members to be easier informed of changes to help avoid conflicts and infrastructure tools which assist in improving interoperability by streamlining data and control integration. Notice the heavy emphasis on tools. Due to the sheer volume of available tools, there is no shortage of options. However, Gupta and Fernandez also found through experience that “the choices of how to collaborate is usually an ad hoc decision based on current practices, the available collaboration mechanisms and sometimes, client-related constraints. Their investigations reveal few examples in which software teams have planned a collaboration strategy based on conscious analysis of collaboration needs, constraints in the project, tools availability, etc.” This lack of emphasis on tool analysis before beginning projects can heavily impact productivity and lead to frustration and a decline in morale \cite{16}.   
The open-source software development space is one place to look for difficulties in software collaboration and how to combat them. Braunschweig and Seaman took an in-depth look at open-source software. They determined that developers must have a shared understanding of the goals and plans of the project to avoid complications and misunderstandings. To evaluate increasing shared knowledge, they attempted to quantify four attributes of communication: Synchrony, Proximity, Proportionality, and Maturity. They found that highly mature contact was typically associated with “a high degree of structure and standardization. Standards, tools, explicit rules, and mentoring of new developers help ensure that work is performed following consistent processes, without much ongoing discussion of those processes.” This finding also supported their other results, indicating that having standard processes and tools helps all aspects of communication and remote software development \cite{17}. Reaching this maturity is the tricky part and is impacted by the factors previously discussed, such as team trust, team dynamic. The ability to select a proper toolset and standards for the team \cite{a14}.

\section{The role of AI in shifting companies to remote working}
As mentioned earlier, in the new era, they are moving towards working remotely, while this method has many challenges that must be taken into account. With the advancement of information and communication technology, new technologies such as AI can play a complementary role in remote working and reduce existing challenges \cite{18}. AI leads processes in a company to be more intelligent and more automatic \cite{19} and, in this way, can assist remote working by improving efficiency \cite{20}. In the following, we will review some applications of AI in remote first companies. In a remote-first company, meetings are an essential part of work. These meetings can be between employees of a team with customers, employees of a team with each other,  and also with other teams of the company. During these meetings, critical information is exchanged. So, meetings should be high-quality in terms of sound, resolution, etc., so employees can provide and receive important information. The settings of meeting software may be very frustrating for IT leaders. If something goes wrong in the setting, the meeting should be disrupted to set them again, which is time-consuming. Recently, AI-driven software can assist IT leaders in their job by setting light, resolution, sound, giving access to the right member, etc. \cite{21}. In addition, during the meeting, AI can collect and transfer information automatically \cite{22}. Another critical challenge in remote working is communicating across teams at different levels and following up with them \cite{23}. In an AI-driven application, employees can enter their requests into an AI bot, and the AI bot can communicate with the right person and ask for their task status \cite{24}. This can be time-consuming and confusing without AI. Moreover, in remote work, employees usually struggle with many technical problems, and because they don't access a rapid IT expert, they have to solve these problems by themselves. In this field, AI can play a useful role. Companies usually use AI-driven agents for online troubleshooting. In this way, when remote employees have a technical problem, contact an AI bot and enter their problem. The AI bot provides suggestions, and a step-by-step solution to the problem \cite{25}. These examples show that AI offers different remote working services and can help companies to bridge the gap of remote working.
\section{	CONCLUSION}
The remote-first working environment is still an open discussion.  Since we have the human factor involved in the discussion, it is almost impossible to say whether remote-first is the best idea.  As we mentioned in section III, every individual has their preferences, which is one of the biggest reasons this discussion is open.  The remote-first work environment has positive and negative sides, no matter how experienced or seasoned employees are.  In terms of productivity, on the other hand, based on the number of pushes, pull requests, code reviews, and commented issues on remote repositories made by software engineers, they kept their pace in a remote environment compared to the office, which proves us well organized remote teams can still be as productive as they were on-site.  [8] While the article explains productivity through GitHub pushes, pull requests, and other measurements, there are many other factors to look at when it comes to productivity.  Based on the findings in this paper, overall productivity has declined due to the struggles associated with working with a remote team.  One of the biggest challenges in remote-first work environments is harmony within the remote team.  The influential factors which make it challenging are team trust and dynamics.  
As natural behavior of us human beings, we tend not to trust people we just met at the beginning.  Based on the nature of humans, this process becomes even more challenging when people/teams have their first meeting via online platforms where they cannot see each other’s facial impressions or read their body language as clearly as they do in physical meetings.  Choosing the proper toolset to improve communication and collaboration within the team companies can ease this challenge.

\bibliographystyle{IEEEtran}
\bibliography{References}

\end{document}